\documentclass[12pt]{article}
\begin{document}
\begin{flushright}
UNB Technical Report: 03-01 \\ CECS-PHY-03/01
\end{flushright}
\vspace{0.4cm}

\begin{center}

{\Large {\bf A finite action for three dimensional gravity with a minimally coupled 
scalar field}} 
\vskip 0.8truecm 
{\large Jack Gegenberg$^{\,\S}$, Cristi\'{a}n Mart\'{\i }nez$^{\,\ddag}$, and Ricardo Troncoso$^{\,\ddag}$}
\vskip 0.2cm 
 \small{
$^{\,\S}${\it Department of Mathematics and Statistics and Department of Physics,\\
University of New Brunswick, Fredericton, NB Canada E3B 5A3}\\
$^{\ddag}${{\it Centro de Estudios
Cient\'{\i}ficos (CECS), Casilla 1469, Valdivia, Chile} }  } \\
\small 
{e-mail: {\tt \footnotesize{ lenin@math.unb.ca, martinez@cecs.cl, ratron@cecs.cl } } }
\vskip 0.5truecm

\end{center}

\vskip 0.5cm

\begin{abstract}

Three-dimensional gravity with a minimally coupled self-interacting scalar
is considered. 
The fall-off of the fields at infinity is assumed to be
slower than that of a localized distribution of matter, so that the
asymptotic symmetry group is the conformal group.

The counterterm Lagrangian needed to render the action finite is found
by demanding that the action attain an extremum for the boundary conditions
implied by the above fall-off of the fields at infinity. These counterterms 
explicitly depend on the scalar field. As a consequence, the Brown-York
stress-energy tensor acquires a non trivial contribution from the matter
sector.

Static circularly symmetric solutions with a regular scalar field are
explored for a one-parameter family of potentials. Their masses are computed 
via the Brown-York quasilocal stress-energy tensor, and they coincide with
the values obtained from the Hamiltonian approach. The thermal behavior,
including the transition between different configurations, is analyzed, and
it is found that the scalar black hole can decay into the BTZ solution
irrespective of the horizon radius.

It is also shown that the AdS/CFT correspondence yields the same central
charge as for pure gravity.
\end{abstract}

\section{Introduction}

The asymptotic behavior of gravity with a negative cosmological constant has
been extensively explored since the 1980's, initially in the context
of symmetries and conserved charges \cite{Abbott:1981ff}-\cite
{Barnich:2003xg}, and now in relation with the AdS/CFT correspondence 
\cite{Maldacena:1997re,Gubser:1998bc,Witten:1998qj}. The usual assumption is
that matter fields fall off sufficiently fast to insure that conserved
charges can be written as surface integrals involving only the metric and
its derivatives. Here we deal with a case where the matter fields
drop off so slowly in the asymptotic region, that they add a nontrivial
contribution to the conserved charges, as well as to the Euclidean action.
This issue is addressed for three-dimensional gravity with a minimally
coupled self-interacting real scalar field. This theory admits interesting
asymptotically AdS solutions including black holes \cite{HMTZ} and a sort of
degenerate ground state, both with non-trivial regular scalar fields.  
To distinguish those black holes with a non-trivial scalar field from BTZ 
black holes (which  are solutions with the scalar field constant) we 
call the former `scalar black holes'.  With 
the exception of 
the BTZ geometries, the solutions have a slower than expected fall-off
to AdS in the asymptotic region. The scalar potential
is constructed so that this weaker fall-off is preserved by the action
of the Virasoro group generated by the asymptotic Killing vector fields.

Since the scalar field behaves as $O(r^{-1/2})$,  it necessarily contributes to the
action and its variations in the asymptotic region, and the counterterms
of  \cite{Henningson:1998gx}  will not yield a finite action or charges \cite
{Balasubramanian-Kraus} in this
case. In the next section, we obtain suitable counterterms, which depend
explicitly on the scalar field, from the requirement
that the action must be functionally differentiable for both metric
and scalar fields which obey these weaker fall-off conditions. This means that the
quasilocal stress-energy tensor defined in Ref. \cite{Brown-York} also
acquires a contribution coming from the matter field. In Section \ref{3}, we
will first briefly review the asymptotic conditions of Ref. \cite{HMTZ}, and
we will display static circularly symmetric solutions with a regular
scalar field for a one-parameter family of potentials, and compute the mass 
using the Brown-York stress-energy tensor. In Section \ref{4}, we will
calculate the action for the Wick-rotated solutions, and use this to discuss
some aspects of the thermodynamics of the solutions. It turns out that there
is a nonvanishing probability for the decay of a scalar black hole into the
BTZ black hole. The central charge is computed via the AdS/CFT
correspondence, yielding the same value as one would obtain for pure gravity \cite
{Brown-Henneaux}.

\section{Action, counterterms and quasilocal stress-energy tensor}

\subsection{Asymptotic fall-off conditions}

The asymptotic behavior of three-dimensional pure gravity with a negative
cosmological constant is described by the Brown-Henneaux boundary conditions 
\cite{Brown-Henneaux}, which are left invariant under a symmetry group
generated by the following asymptotic Killing vectors 
\begin{eqnarray}
\eta ^{t} &=&l\left[ T^{+}+T^{-}+\frac{l^{2}}{2r^{2}}(\partial
_{+}^{2}T^{+}+\partial _{-}^{2}T^{-})\right] +O(r^{-4})\;,  \nonumber \\
\eta ^{r} &=&-r(\partial _{+}T^{+}+\partial _{-}T^{-})+O(r^{-1})\;,
\label{AsymptKillingVectors} \\
\eta ^{\varphi } &=&T^{+}-T^{-}-\frac{l^{2}}{2r^{2}}(\partial
_{+}^{2}T^{+}-\partial _{-}^{2}T^{-})+O(r^{-4})\;,  \nonumber
\end{eqnarray}
where $T^{+}(x^{+})$ and $T^{-}(x^{-})$ generate two independent copies of
the Virasoro algebra and $x^{\pm }=t/l\pm \varphi $.

These conditions hold also for localized matter fields which fall-off
sufficiently fast at infinity, so as to give no contributions to the surface
integrals defining the generators of the asymptotic symmetries. With these
assumptions, the charges that generate the asymptotic symmetries involve
only the metric and its derivatives, and their algebra corresponds to be a
central extension of the asymptotic symmetry algebra, where the central
charge is given by 
\begin{equation}
c=\frac{3l}{2G}\,.  \label{Central Charge}
\end{equation}

However, there are instances in which the matter fields modify the
asymptotic behavior of the metric. A well-known example is the electrically
charged black hole, where the metric has a logarithmic divergence
 \cite{BTZ, MTZ, Ashtekar:2002qc}. 
In those cases, there is a possibility of having divergent
contributions coming both from the gravitational and matter actions. In
these situations, the asymptotic conditions must be such that the sum of
both contributions converges.

The case in which the fall-off of the fields at infinity is slower than that
of a localized distribution of matter was analyzed in Ref. \cite{HMTZ}. The
matter sector was assumed to be given by a single self-interacting scalar
field minimally coupled to three dimensional gravity, with the action 
\begin{equation}
I_{Bulk}[g,\phi ]=\frac{1}{\pi G}\int d^{3}x\sqrt{-g}\left[ \frac{R}{16}-%
\frac{1}{2}(\nabla \phi )^{2}-V(\phi )\right] \;.  \label{Action Bulk}
\end{equation}
Black hole solutions with a nontrivial scalar field were found for a
one-parameter family of potentials $V(\phi )$, whose asymptotic behavior
belongs to the following class 
\begin{equation}
\phi =\frac{\chi }{r^{1/2}}-\alpha \frac{\chi ^{3}}{r^{3/2}}+O(r^{-5/2})
\label{AsymptPhi}
\end{equation}
\begin{equation}
\begin{array}{lll}
g_{rr}=\displaystyle \frac{l^{2}}{r^{2}}-\frac{4l^{2}\chi ^{2}}{r^{3}}%
+O(r^{-4}) &  & \displaystyle g_{tt}=-\frac{r^{2}}{l^{2}}+O(1) \\[2mm]
g_{tr}=O(r^{-2}) &  & g_{\varphi \varphi }=r^{2}+O(1) \\[1mm]
g_{\varphi r}=O(r^{-2}) &  & g_{t\varphi }=O(1)
\end{array}
\label{AsymptGeneral}
\end{equation}
where $\chi =\chi (t,\varphi )$, and $\alpha $ is an arbitrary constant.
Note that the asymptotic behavior of $g_{rM}$ has a slower fall-off than that 
discussed 
by Brown and Henneaux\footnote{%
Henceforth, capital latin indices stand for three-dimensional spacetimes
coordinates, and greek indices label the coordinates at the boundary.}.
Consistency of these relaxed asymptotic boundary conditions with the field
equations is sufficient to fix the potential $V(\phi )$ to be of the form 
\begin{equation}
V(\phi )=-\frac{1}{8l^{2}}-\frac{3}{8l^{2}}\phi ^{2}-\frac{1}{2l^{2}}\phi
^{4}+\phi ^{6}U(\phi )\;,  \label{Generic Potential}
\end{equation}
where $U(\phi ^{2})$\ could be any function which is smooth in a
neighborhood of $\phi =0$. In spite of the fact that $V(\phi )$\ could even
be unbounded from below, this potential satisfies the conditions that
guarantee the perturbative stability of AdS space \cite{B-F}, \cite{M-T}.

Remarkably, it was found that this set of conditions is also left invariant
under the Virasoro algebra generated by the asymptotic Killing vectors (\ref
{AsymptKillingVectors}). Furthermore, using the Regge-Teitelboim approach 
\cite{Regge:1974zd}, it was found that the generators of the asymptotic
symmetries acquire a contribution from the scalar field\footnote{%
Eq.(\ref{Canonical Charges}) is a slightly improved version of the
expression found in Ref. \cite{HMTZ}, because it does not depend on the
parameter $\alpha $ appearing in Eq.(\ref{AsymptPhi}).} 
\begin{eqnarray}
Q(\xi ) &=&\frac{1}{16\pi G}\int d\varphi \left\{ \frac{\xi ^{\bot }}{lr}%
\left( (g_{\varphi \varphi }-r^{2})-2r^{2}(lg^{-1/2}-1)\right)+2\xi ^{\varphi }\pi _{\varphi }^{r} \right. 
\nonumber \\
&&\left.  +
 \xi ^{\bot } \frac{2r}{l}\left[ \phi ^{2}-2l\frac{\phi \partial _{r}\phi }{%
\sqrt{g_{rr}}}\right]  \right\}
\,,  \label{Canonical Charges}
\end{eqnarray}
and the algebra of these canonical generators have the standard central
extension given by Eq. (\ref{Central Charge})

In the next subsection, we use the background independent 
method of \cite{Henningson:1998gx} to find the counterterm Lagrangian needed to render the
action finite (see also \cite{Balasubramanian-Kraus}).  
In contrast to the
case of a localized distribution of matter, it is shown that the correction
terms acquire contributions depending explicitly on the scalar field. This
allows us to construct an alternative to Eq. (\ref{Canonical Charges}) which we 
obtain from the Brown-York stress-energy tensor.

\subsection{Counterterm action}

In analogy with the counterterm prescription  in Refs. \cite
{Henningson:1998gx, Balasubramanian-Kraus}, we consider the following action 
\begin{equation}
I=\frac{1}{\pi G}\int_{M}d^{3}x\sqrt{-g}\left( \frac{R}{16}-\frac{1}{2}%
(\nabla \phi )^{2}-V(\phi )\right) +\frac{1}{8\pi G}\int_{\partial M}d^{2}x%
\sqrt{-\gamma }\Theta +I_{ct}[\gamma ,\phi ]\;,  \label{ITotal}
\end{equation}
where the boundary term containing the trace of the extrinsic curvature $%
\Theta $ is required to fix Dirichlet conditions for the metric. Owing to
the asymptotic behavior of the fields, the counterterm action $I_{ct}$ is
assumed to depend not only on the boundary metric $\gamma _{\mu \nu ,}$, but
also on the scalar field.

The strategy for obtaining the counterterm action will be to require that the
action (\ref{ITotal}) should have an extremum for the solutions satisfying
the asymptotic conditions (\ref{AsymptPhi}) and (\ref{AsymptGeneral}). It
turns out that this approach also ensures the convergence of the action.

The variation of (\ref{ITotal}) is given by 
\begin{equation}
\delta I=\int_{M}E_{I}\delta \Phi ^{I}+\int_{\partial M}d^{2}x\;\pi ^{\mu
\nu }\delta \gamma _{\mu \nu }-\frac{1}{\pi G}\int_{\partial M}d^{2}x\sqrt{%
-\gamma }\hat{n}^{M}\partial _{M}\phi \delta \phi +\delta I_{ct}\;,
\label{Delta I0}
\end{equation}
where $\Phi ^{I}:=\left\{ g_{MN},\phi \right\} $ are the dynamical fields
and $E_{I}$ are the corresponding equations of motion. The spacetime metric $%
g_{MN}$ has been decomposed in a radial ADM foliation as 
\begin{equation}
ds^{2}=N^{2}dr^{2}+\gamma _{\mu \nu }(dx^{\mu }+N^{\mu }dr)(dx^{\nu }+N^{\nu
}dr)\;,  \label{ADM-Radial}
\end{equation}
and the boundary momenta are 
\begin{equation}
\pi ^{\mu \nu }:=\frac{1}{16\pi G}\sqrt{-\gamma }\left( \Theta \gamma ^{\mu
\nu }-\Theta ^{\mu \nu }\right) \;.  \label{Pi}
\end{equation}
Here the extrinsic curvature $\Theta ^{\mu \nu }$, defined through the
covariant derivative of the outward pointing normal vector $\hat{n}_{M}=(0,%
1/\sqrt{g^{rr}},0)$ to the boundary $\partial M$, is
\begin{equation}
\Theta ^{\mu \nu }:=\frac{1}{2}\left( \nabla ^{\mu }\hat{n}^{\nu }+\nabla
^{\nu }\hat{n}^{\mu }\right) \;.  \label{Theta}
\end{equation}
Hence the action attains an extremum provided the variation of the
counterterm action satisfies 
\[
\delta I_{ct}=-\int_{\partial M}d^{2}x\;\pi ^{\mu \nu }\delta \gamma _{\mu
\nu }\;+\frac{1}{\pi G}\int_{\partial M}d^{2}x\sqrt{-\gamma }\hat{n}%
^{M}\partial _{M}\phi \delta \phi \;,
\]
which, by virtue of the asymptotic conditions (\ref{AsymptPhi}) and (\ref
{AsymptGeneral}), becomes
\begin{equation}
\delta I_{ct}=-\frac{1}{16\pi G}\int_{\partial M}d^{2}x\;\left( \frac{1}{%
l^{2}}\delta h_{\varphi \varphi }-\delta h_{tt}\right) -\frac{1}{4\pi Gl^{2}}%
\int_{\partial M}d^{2}x\left( r\delta \chi ^{2}+(1+3\alpha )\delta \chi
^{4}\right) \;.  \label{DeltaIct}
\end{equation}
Here $h_{\mu \nu }(t,\varphi )$ denotes the deviation from the AdS
asymptotic metric. The first term at the right hand side of (\ref{DeltaIct})
corresponds to the variation of the volume of $\partial M$ 
\[
\delta \left[ \int_{\partial M}d^{2}x\;\sqrt{-\gamma }\right] =\frac{l}{2}%
\int_{\partial M}d^{2}x\;\left( \frac{1}{l^{2}}\delta h_{\varphi \varphi
}-\delta h_{tt}\right) \;,
\]
and using the asymptotic form of the scalar field, it is simple to see that
the second term in (\ref{DeltaIct}) corresponds to the variation of a
covariant expression that reads 
\[
\frac{1}{8\pi Gl}\int_{\partial M}d^{2}x\;\sqrt{-\gamma }\left( 2l\;\phi 
\hat{\eta}^{M}\partial _{M}\phi -\phi ^{2}\right) \;,
\]
which depends on the boundary metric as well as on the scalar field.

Therefore, the counterterm action can be written as 
\begin{eqnarray}
I_{ct} &=&I_{ct}^{G}+I_{ct}^{\phi }  \nonumber \\ \label{Ict}
&=&\frac{1}{8\pi Gl}\left[ -\int_{\partial M}d^{2}x\sqrt{-\gamma }%
+\int_{\partial M}d^{2}x\sqrt{-\gamma }\left( 2l\;\phi \hat{\eta}%
^{M}\partial _{M}\phi -\phi ^{2}\right) \right],
\end{eqnarray}
where $I_{ct}^{G}$ is the counterterm required for asymptotically AdS
spacetimes in the sense of Brown and Henneaux, and $I_{ct}^{\phi }$ is
required to cancel the variation coming from the kinetic term in the bulk,
provided the relaxed asymptotic conditions (\ref{AsymptPhi}) and (\ref
{AsymptGeneral}) are imposed.

The presence of $I_{ct}^{\phi }$ implies that the surface integrals defining
the conserved charges acquire a nontrivial contribution coming from the
matter sector. This will be explicitly demonstrated in the next subsection, where we
employ the Brown-York approach.

\subsection{Quasilocal stress-energy tensor}

The quasilocal stress-energy tensor $T^{\mu \nu }$ associated with a region
of the spacetime that is bounded by a surface with metric $\gamma _{\mu \nu }
$ is given by 
\[
T^{\mu \nu }=\frac{2}{\sqrt{-\gamma }}\frac{\delta I}{\delta \gamma _{\mu
\nu }}\;.
\]
Considering a radial ADM foliation as in Eq. (\ref{ADM-Radial}), the
variation of the action on-shell is 
\[
\frac{\delta I}{\delta \gamma _{\mu \nu }}=\int_{\partial M}d^{2}x\;\pi
^{\mu \nu }+\frac{\delta I_{ct}}{\delta \gamma _{\mu \nu }}
\]
where $\pi ^{\mu \nu }$ is given by (\ref{Pi}). Hence, using Eq. (\ref{Pi}),
the stress-energy tensor reads:
$$
T^{\mu \nu }=\frac{1}{8\pi G}\left(\Theta \gamma ^{\mu \nu }-\Theta ^{\mu
\nu }\right) +\frac{2}{\sqrt{-\gamma }}\frac{\delta I_{ct}}{\delta \gamma
_{\mu \nu }}\;,
$$
which by virtue of $I_{ct}$ in Eq. (\ref{Ict}), can be written as
\begin{eqnarray}
T^{\mu \nu } &=&T_{G}^{\mu \nu }+T_{\phi }^{\mu \nu }  \nonumber \\
&=&\frac{1}{8\pi G}\left( \Theta \gamma ^{\mu \nu }-\Theta ^{\mu \nu }-\frac{%
1}{l}\gamma ^{\mu \nu }\right) +\frac{1}{8\pi Gl}\gamma ^{\mu \nu }\left(
2l\;\phi \hat{\eta}^{M}\partial _{M}\phi -\phi ^{2}\right).  \label{Tmn}
\end{eqnarray}

The conserved charges can be constructed by choosing an ADM foliation at $%
\partial M$ with spacelike surfaces $\Sigma $, so that 
\[
\gamma _{\mu \nu }dx^{\mu }dx^{\nu }=-N_{\Sigma }^{2}dt^{2}+\sigma
(d\varphi+N_{\Sigma }^{\varphi}dt)^2\;.
\]
Hence
\[
Q_{BY}(\xi )=\int_{\Sigma }dx\;\sqrt{\sigma }u^{\mu }\xi ^{\nu }T_{\mu \nu
}\;.
\]
Here $u^{\mu }$ is the timelike unit normal to $\Sigma $, and $\xi ^{\mu }$
is a Killing vector of the boundary. Thus, choosing $\xi ^{\mu }=N_{\Sigma
}u^{\mu }$, the mass is written as a surface integral, 
\begin{eqnarray}
M &=&\int_{\Sigma }dx\;\sqrt{\sigma }N_{\Sigma }u^{\mu }u^{\nu }(T_{\mu \nu
}^{G}+T_{\mu \nu }^{\phi })  \nonumber \\
&=&M_{G}+M_{\phi }\;.  \label{Mg+Mphi}
\end{eqnarray}
Note that the mass acquires a nontrivial contribution from the matter
sector.

In what follows, the previous formalism is tested for some exact solutions
possessing the asymptotic behavior given by (\ref{AsymptPhi}) and (\ref
{AsymptGeneral}) for a one-parameter family of potentials of the form (\ref
{Generic Potential}).

\section{Testing the counterterms with exact solutions} \label{3}

Exact solutions for which the metric and the scalar field satisfy the
asymptotic conditions (\ref{AsymptPhi}) and (\ref{AsymptGeneral}) can be obtained
for a\ particular one-parameter family of potentials of the form \cite{HMTZ} 
\begin{equation}
V_{\nu }(\phi )=-\frac{1}{8l^{2}}\left( \cosh ^{6}\phi +\nu \sinh ^{6}\phi
\right) \;.  \label{The Potential}
\end{equation}
This potential belongs to the class (\ref{Generic Potential}), and different
forms of $U(\phi ^{2})$ are obtained for different values of the
dimensionless parameter $\nu $. This parameter can be interpreted as the
self-interacting coupling constant in the conformal frame \cite{HMTZ}.

\subsection{Black hole with a regular non-constant scalar field}

For the range $\nu >-1$, the potential is unbounded from below and satisfies
the conditions that guarantees the perturbative stability of AdS space \cite
{M-T}. In this case, a static circularly symmetric black hole solution,
dressed with a scalar field which is regular everywhere\footnote{%
The solution for $\nu =0$ was found in the conformal frame in Ref. \cite
{Martinez:1996gn}. Recently, a four dimensional black hole dressed with a
conformally coupled scalar field has been obtained in \cite{Martinez:2002ru}
for a positive cosmological constant.} was found in Ref. \cite{HMTZ}. The
scalar field is given by

\begin{equation}
\phi =\textrm{arctanh}\sqrt{\frac{B}{H(r)+B}}\;,  \label{Scalar}
\end{equation}
where $B$ is a nonnegative integration constant and 
\[
H(r)=\frac{1}{2}\left( r+\sqrt{r^{2}+4Br}\right) \;.
\]
The metric reads 
\begin{equation}
ds^{2}=-\left( \frac{H}{H+B}\right) ^{2}F(r)dt^{2}+\left( \frac{H+B}{H+2B}%
\right) ^{2}\frac{dr^{2}}{F(r)}+r^{2}d\varphi ^{2}\;,  \label{The metric}
\end{equation}
with 
\[
F=\frac{H^{2}}{l^{2}}-(1+\nu )\left( \frac{3B^{2}}{l^{2}}+\frac{2B^{3}}{%
l^{2}H}\right) \;.
\]
The event horizon is located at

\[
r_{+}=B\Theta _{\nu }\,, 
\]
where the constant $\Theta _{\nu }$\textbf{\ }is expressed in terms of $z=1+i%
\sqrt{\nu }$, as 
\begin{equation}
\Theta _{\nu }=2(z\bar{z})^{2/3}\frac{z^{2/3}-\bar{z}^{2/3}}{z-\bar{z}}\;.
\label{Schuster}
\end{equation}
As a function of $\nu $, $\Theta _{\nu }$ is monotonically increasing, and
asymptotically grows as $\sqrt{\nu }$. The causal structure of this geometry
is identical to that of the non-rotating BTZ black hole \cite{BTZ}.

The mass of this black hole can be obtained from the quasilocal
stress-energy tensor (\ref{Tmn}) choosing $u^{\mu }=\frac{1}{N_{\Sigma }}%
\delta _{t}^{\mu }$ in the surface integral (\ref{Mg+Mphi}), which now reads

\begin{eqnarray*}
M &=&M_{G}+M_{\phi }=-\lim_{r\rightarrow \infty }\int_{S^{1}}rd\varphi \;%
\sqrt{-g_{tt}}g_{tt}(T_{G}^{tt}+T_{\phi }^{tt})\;, \\
&=&\frac{1}{8\pi G}\lim_{r\rightarrow \infty }\int_{S^{1}}rd\varphi \;\sqrt{%
-g_{tt}}\left( \left( -\frac{1}{r\sqrt{g_{rr}}}+\frac{1}{l}\right) +\frac{1}{%
l}\left( \phi ^{2}-\frac{2l}{\sqrt{g_{rr}}}\;\phi \partial _{r}\phi \right)
\right) \;.
\end{eqnarray*}
Note that the contribution coming from the matter piece 
\[
M_{\phi }=\frac{1}{2Gl^{2}}\left( Br-B^{2}\right) \;, 
\]
has a linearly divergent term which is exactly cancelled by the one
appearing in $M_{G}$, which is given by 
\[
M_{G}=\frac{1}{2Gl^{2}}\left( -Br+\frac{B^{2}(7+3\nu )}{4}\right) \;, 
\]
and therefore, the black hole mass is 
\begin{equation}
M=\frac{3B^{2}}{8Gl^{2}}(1+\nu )\;.  \label{BH-Mass}
\end{equation}
This result coincides with the expression (\ref{Canonical Charges}) obtained
from the Hamiltonian formalism for $\xi =\partial _{t}$, \textit{i.e.}, $%
Q(\partial _{t})=M$.

\subsection{New solutions}

For the range $\nu \leq 0$, an independent static circularly symmetric
solution with a nontrivial scalar field exists\footnote{%
In the conformal frame \cite{HMTZ}, this solution corresponds to a massless
BTZ black hole dressed with a scalar field given by $\hat{\phi}=\sqrt{\frac{B%
}{r+B\sqrt{-\nu }}}$, which for $\nu <0$ is regular everywhere. For the case 
$\nu =0$ \cite{Natsuume:1999cd, Ayon-Beato:2001sb}, the scalar field
diverges at the origin.}. The scalar field is given by

\begin{equation}
\phi =\textrm{arctanh}\left( \sqrt{\frac{B}{f(r)+B\sqrt{-\nu }}}\right) \;,
\label{PhiNS}
\end{equation}
with

\begin{equation}
f(r)=\frac{1}{2}\left( r-B(\sqrt{-\nu }-1)+\sqrt{\left( r-B\left( \sqrt{-\nu 
}-1\right) \right) ^{2}+4B\sqrt{-\nu }r}\right) \;,  \label{f}
\end{equation}
so that $\alpha $ in Eq. (\ref{AsymptPhi}) differs from $-\frac{2}{3}$,  
the value it attains in the black hole case, but now depends on the parameter $\nu $:
\[
\alpha =-\frac{1+3\sqrt{-\nu }}{6}\;.
\]
The metric is given by

\begin{equation}
ds^{2}=-\frac{r^{2}}{l^{2}}dt^{2}+\frac{l^{2}(f+B(\sqrt{-\nu }-1))^{2}(f+B%
\sqrt{-\nu })^{2}}{f^{2}(f^{2}-B(B-2f)\sqrt{-\nu }-B^{2}\nu )^{2}}%
dr^{2}+r^{2}d\varphi ^{2}\;,  \label{NS}
\end{equation}
with asymptotic behavior of the form (\ref{AsymptGeneral}). The
integration constant $B$ is nonnegative, and for $B=0$, this solution
reduces to the massless BTZ black hole with $\phi =0$. The geometric
behavior of (\ref{NS}) radically varies, depending on the range of the
parameter $\nu $.

\subsubsection{The Nullnut} \label{nullnutsec}

For the range $\nu <-1$, the potential looks like a ``Mexican hat''. In this
case, the scalar field (\ref{PhiNS}) is regular everywhere, and the metric (%
\ref{NS}) possesses a timelike killing vector whose norm vanishes at $r=0$.
Note that, under a suitable time rescaling, the line element around $r=0$
can be written as a massless BTZ black hole with an effective AdS radius $%
\tilde{l}=l\frac{\sqrt{-\nu }-1}{\sqrt{-\nu }}$, 
\[
ds_{r\rightarrow 0}^{2}=-\frac{r^{2}}{\tilde{l}^{2}}dt^{2}+\frac{\tilde{l}%
^{2}}{r^{2}}dr^{2}+r^{2}d\varphi ^{2}\;,
\]
This means that the geometry is smooth, as can be seen from the behavior of
the Ricci scalar near the origin 
\[
R_{r\rightarrow 0}=-\frac{6}{\tilde{l}^{2}}+O(r^{2})\;.
\]
This geometry has been dubbed ``nullnut'' because it has a nut  on the null
curve $r=0$. Its causal structure is the same as for the massless BTZ black
hole, irrespective of the value of the integration constant $B$. As can be
foreseen through the invariance under boosts in the $t-\varphi $ plane, this
solution, independently of the integration constant $B$, has a vanishing
mass. This can be explicitly checked from the quasilocal stress-energy
tensor (\ref{Tmn}) by choosing $u^{\mu }=\frac{1}{N_{\Sigma }}\delta
_{t}^{\mu }$ in the surface integral (\ref{Mg+Mphi}). In this case, the
contribution coming from the matter piece, 
\[
M_{\phi }=\frac{Br}{2l^{2}}+\frac{B^{2}(1-3\sqrt{-\nu })}{4l^{2}}\;,
\]
is exactly cancelled by the contribution of the gravitational sector, that
is, $M_{G}=-M_{\phi }$, yielding zero total mass. Owing to this
fact, these configuration can be regarded as a sort of degenerate ground
state.

For the range $-1\leq \nu \leq 0$, the metric (\ref{NS}) describes a
different geometry, because it has what we consider to be a mild naked
singularity at $r=0$, since its mass vanishes, and it has a finite Euclidean
action.

\section{Thermodynamics from the Euclidean action} \label{4}

Since our action (\ref{ITotal}) has been regularized by the presence of the
counterterm (\ref{Ict}), in the semiclassical approximation the partition
function is determined by the exponential of the Euclidean action, $Z=\exp
(I)$, evaluated on the classical solution, without the need of a background
substraction.

Hence, the thermodynamics for the scalar black hole (\ref{Scalar},\ref{The
metric}) can be read from the Euclidean action evaluated on the Wick-rotated
solution.

On shell, the bulk term reduces to 
\begin{eqnarray*}
I_{Bulk} &=&\frac{2}{\pi G}\int_{M}d^{3}x\sqrt{g}V(\phi ) \\
&=&-\frac{\beta }{2Gl^{2}}\lim_{r_{0}\rightarrow \infty
}\int_{r_{+}}^{r_{0}}dr\frac{r\left[ (H+B)^{3}+\nu B^{3}\right] }{H^{2}(H+2B)%
} \\
&=&-\frac{\beta }{4Gl^{2}}[r_{0}^{2}+2Br_{0}-B^{2}(5+3\nu )]\;,
\end{eqnarray*}
and, when the cut-off $r_{0}\rightarrow \infty $, the surface terms read

\begin{eqnarray*}
\frac{1}{8\pi G}\int_{\partial M}d^{2}x\sqrt{\gamma }\Theta &=&\frac{\beta }{%
4Gl^{2}}\left[ 2r_{0}^{2}+4Br_{0}-B^{2}(7+3\nu )\right] \;, \\
\frac{1}{8\pi G}\int_{\partial M}d^{2}x\frac{\sqrt{\gamma }}{l} &=&\frac{%
\beta }{4Gl^{2}}\left[ r_{0}^{2}-\frac{3}{2}B^{2}(1+\nu )\right] \;, \\
\frac{1}{8\pi Gl}\int_{\partial M}d^{2}x\;\sqrt{-\gamma }\left( 2l\;\phi 
\hat{\eta}^{M}\partial _{M}\phi -\phi ^{2}\right) &=&\frac{\beta }{4Gl^{2}}%
\left[ 2B^{2}-2Br_{0}\right] \;.
\end{eqnarray*}
As the scalar black hole solution, described by (\ref{Scalar})
and (\ref{The metric}), has an inverse Hawking temperature given by 
\begin{equation}
\beta =\frac{2\pi l^{2}}{3B}\frac{\Theta _{\nu }}{(1+\nu )}\;,
\label{Temperature}
\end{equation}
the Euclidean action is 
\begin{equation}
I=-\beta F=\frac{\pi \Theta _{\nu }}{4G}B\;,  \label{Ie}
\end{equation}
and consequently, the counterterm prescription reproduces the expected
thermodynamic expression for the free energy. Indeed, the mass in Eq.(\ref
{BH-Mass}) and the entropy are recovered from (\ref{Ie}) through 
\begin{eqnarray*}
M &=&-\frac{\partial I}{\partial \beta }=\frac{3B^{2}}{8Gl^{2}}(1+\nu )\;, \\
S &=&(1-\beta \frac{\partial }{\partial \beta })I=\frac{\pi r_{+}}{2G}=\frac{%
A}{4G}\;.
\end{eqnarray*}

An analogous computation shows that, for the nullnut solution given by (\ref{PhiNS})
and (\ref{NS})  with $\nu <-1$, the Euclidean action vanishes. This
is consistent with the fact that the nullnut has vanishing mass and
temperature, as well as a null entropy.

\subsection{Thermal Decay}

Note that the specific heat of the scalar black hole is given by $C=\partial
M/\partial T=\frac{\pi }{2}r_{+}$, which is always positive. This means that
the scalar black hole can always reach thermal equilibrium with a heat bath.
However, for a fixed temperature, apart from the scalar black hole (SBH), a
BTZ black hole with a vanishing scalar field can also be at equilibrium with
the heat bath. This raises the question of whether the scalar black
hole could decay into the BTZ black hole. To examine this question, one
needs to evaluate the difference between their respective free energies.

As both geometries approach AdS at infinity, the matching of the
temperatures leads to the following relationship between both horizon radii:
\[
r_{+}^{SBH}=\frac{\Theta _{\nu }^{2}}{3(1+\nu )}r_{+}^{BTZ}\;.
\]
The action for the scalar black hole is given by (\ref{Ie}), and the action
for BTZ can be readily obtained in the same way: $I_{BTZ}=\pi
r_{+}^{_{BTZ}}/{4G}$. Therefore their difference 
\[
I_{SBH}-I_{BTZ}=\frac{\pi }{4G}\left[ \Theta _{\nu }-\frac{3(1+\nu )}{\Theta
_{\nu }}\right] B\;,
\]
which is always negative since $\Theta _{\nu }-\frac{3(1+\nu )}{\Theta _{\nu
}}<0$ for the range $\nu >-1$. This means that there is a nonvanishing
probability for the decay of a scalar black hole into the BTZ black hole,
induced by the thermal fluctuations, irrespective of the value of the
horizon radius. Furthermore, since

\[
\frac{M_{SBH}}{M_{BTZ}}=\frac{S_{SBH}}{S_{BTZ}}=\frac{\Theta _{\nu }^{2}}{%
3(1+\nu )}<1\;,
\]
in the decay process, the scalar black hole absorbs energy from the thermal
bath, thus increasing its horizon radius, and consequently its entropy. This
suggests that in this process, the scalar field is, in some sense, absorbed
by the black hole.

Analogously, since the nullnut solution, given by (\ref{PhiNS}) and (\ref{NS}%
), has vanishing temperature and action, there is probability of decaying
into a BTZ black hole. Hence, in a similar way, the nullnut would be able to
absorb its own scalar field.

\subsection{Central charge}

Note that, even though the fall-off of the fields at infinity, given by 
(\ref{AsymptPhi}) and (\ref{AsymptGeneral}), is slower than that of a localized
distribution of matter, this set of conditions is also left invariant under
the Virasoro algebra generated by (\ref{AsymptKillingVectors}). Moreover,
although the expression for canonical charges differs form the one found in 
\cite{Brown-Henneaux}, it was shown in \cite{HMTZ} that their algebra is
identical, i.e., two copies of the Virasoro algebra with exactly the same
central extension. This follows from the results of Ref. \cite
{Brown:ed}, which states that the bracket of two charges provides a
realization of the asymptotic symmetry algebra with a possible central
extension. For the class of potentials which are consistent with the
modified asymptotic behavior (\ref{AsymptPhi}) and (\ref{AsymptGeneral}), the
massless BTZ black hole with vanishing scalar field corresponds to the
ground state. Thus, the central charge can be determined by computing the
variation of the charges on the vacuum. It is simple to check that the same
result is obtained for the nullnut solution given in Sec. \ref{nullnutsec}.

The purpose of this section is to show that the central charge can also be
obtained in the context of the AdS/CFT correspondence.  The latter associates 
the Brown-York quasilocal stress-energy tensor $T^{\mu \nu }$
with the stress-energy tensor of a conformal field theory
living at the spacetime boundary. For the three-dimensional cases considered
here, the boundary metric is conformal to a flat cylinder, i.e., $%
ds^{2}=-\frac{dt^{2}}{l^{2}}+d\varphi ^{2}=-dx^{+}dx^{-}$, with $x^{\pm }=%
\frac{t}{l}\pm \varphi $. It is well known that under diffeomorphisms of the
form 
\begin{eqnarray*}
x^{+} &\rightarrow &x^{+}-\xi ^{+}(x^{+})\;, \\
x^{-} &\rightarrow &x^{-}-\xi ^{-}(x^{-})\;,
\end{eqnarray*}
the $T_{++}$ component of the stress-energy tensor for a two-dimensional CFT
transforms as 
\begin{equation}
\delta _{\xi }T_{++}=(2\partial _{+}\xi ^{+}T_{++}+\xi ^{+}\partial
_{+}T_{++})-\frac{c}{24\pi }\partial _{+}^{3}\xi ^{+}\;,  \label{DeltaTmnCFT}
\end{equation}
where the last term is the quantum anomaly. Hence, the central charge $c$
can be obtained by identifying the Brown-York tensor in Eq. (\ref{Tmn}) with
that of the dual CFT. Indeed, making the variation of (\ref{Tmn}) using the
asymptotic Killing vectors given by 
\begin{eqnarray*}
x^{+} &\rightarrow &x^{+}-\xi ^{+}-\frac{l^{2}}{2r^{2}}\partial _{-}^{2}\xi
^{-}\;, \\
x^{-} &\rightarrow &x^{-}-\xi ^{-}-\frac{l^{2}}{2r^{2}}\partial _{+}^{2}\xi
^{+}\;, \\
r &\rightarrow &r+\frac{r}{2}(\partial _{+}\xi ^{+}+\partial _{-}\xi ^{-})\;,
\end{eqnarray*}
and evaluating the expression for the ground state (for which $T_{\mu \nu }$
vanishes) one obtains 
\begin{equation}
\delta _{\xi }T_{++}=-\frac{l}{16\pi G}\partial _{+}^{3}\xi ^{+}\;.
\label{DeltaTmnBY}
\end{equation}
Comparison of (\ref{DeltaTmnBY}) with (\ref{DeltaTmnCFT}) yields the same
expression as one obtains for pure gravity with a negative cosmological constant, 
\[
c=\frac{3l}{2G}\;, 
\]
and hence, the Brown-Henneaux central charge is recovered from the AdS/CFT\
correspondence. It is simple to check that the counterterms explicitly containing
the scalar field do not contribute to the variation of the
quasilocal stress-energy tensor, and also that the same result would hold
if we had chosen the nullnut solution as the ground state.

In the case of pure gravity with the standard asymptotic conditions, the 
central charge was first obtained through the AdS/CFT correspondence by
Henningson and Skenderis \cite{Henningson:1998gx}.

As was noted in \cite{HMTZ}, it is worth mentioning that if one naively
follows the Cardy-Strominger approach \cite{Strominger:1997eq}, for the
black hole solution given by (\ref{Scalar}) and (\ref{The metric}), a value
for the entropy that always exceeds the semiclassical result is obtained,
thus violating in this way the holographic bound \cite{'tHooft:gx,Susskind:1994vu}.

\section{Discussion and comments}

A finite action for three-dimensional gravity with a minimally coupled
self-interacting scalar field has been constructed. Since the fall-off of
the fields at the asymptotic region was assumed to be slower than that of a
localized distribution of matter, the counterterm Lagrangian needed to
render the action finite acquires contributions depending explicitly on the
scalar field. This means that the quasilocal conserved charges also acquires
a non trivial contribution coming from the matter sector. This fact is in
agreement with results obtained via the Hamiltonian \cite{HMTZ} as well
as covariant methods \cite{Barnich:2002pi}.

The same central charge as for pure gravity was found by means of the
AdS/CFT correspondence. The required counterterms were found by demanding that the 
action attain an extremum for the boundary conditions mentioned above,
and this also insures the convergence of the action. A treatment of this
method in a more general settings will appear elsewhere.

The asymptotic conditions considered here correspond to exact solutions
including black holes. Had we assumed different asymptotic behavior for the
fields, different counterterms would be found. This problem has also been explored
 in three and higher dimensions \cite{Nojiri:2000kh, deHaro:2000xn, Berg:2002hy}.

The possibility of a phase transition between black holes and solutions with
naked singularities \cite{Das:2001rk} in three dimensions, as well as its
relevance for the AdS/CFT correspondence have been recently explored
 in Ref. \cite{Das:2001wu}. A more general discussion of the problem of the
thermodynamical properties of naked singularities in this model is currently
under study.

\vskip 0.5cm \textbf{\large Acknowledgments}

The authors are grateful to Juan Cris\'ostomo, Marc Henneaux and Jorge Zanelli for
helpful remarks, and also to Kostas Skenderis and Jacek Wi\'sniewski for 
pointing out to us some references. This work is partially funded by grants 1010446, 1010449,
1010450, 1020629, 7010446, 7010450 from FONDECYT. Institutional support
 to the Centro de Estudios Cient\'{\i}ficos (CECS) from Empresas 
CMPC is gratefully acknowledged.  CECS is a Millennium Science 
Institute and is funded in part by grants from Fundaci\'{o}n Andes and 
the Tinker Foundation.

\end{document}